\documentclass[12pt]{article}

\usepackage[utf8]{inputenc}
\usepackage{epsfig}
\usepackage{amssymb}
\usepackage{graphicx}
\usepackage{wrapfig}
\usepackage{framed}
\usepackage{tabularx}
\usepackage{float}
\usepackage{svg}
\usepackage{comment}
\usepackage{url}
\usepackage{hyperref}

\begin{document}

\title{Latent Space Analysis of VAE and Intro-VAE applied to 3-dimensional MR Brain Volumes of Multiple Sclerosis, Leukoencephalopathy, and Healthy Patients}

\author{
    Christopher Vogelsanger \\ 
    cvogelsa@student.ethz.ch
    \and
    Christian Federau \\ 
    federau@biomed.ee.ethz.ch
}
\date{}

\maketitle

\begin{abstract}
\noindent
Multiple Sclerosis (MS) and microvascular leukoencephalopathy are two distinct neurological conditions, the first caused by focal autoimmune inflammation in the central nervous system, the second caused by chronic white matter damage from atherosclerotic microvascular disease. Both conditions lead to signal anomalies on Fluid Attenuated Inversion Recovery (FLAIR) magnetic resonance (MR) images, which can be distinguished by an expert neuroradiologist, but which can look very similar to the untrained eye as well as in the early stage of both diseases. In this paper, we attempt to train a 3-dimensional deep neural network to learn the specific features of both diseases in an unsupervised manner. For this manner, in a first step we train a generative neural network to create artificial MR images of both conditions with approximate explicit density, using a mixed dataset of multiple sclerosis, leukoencephalopathy and healthy patients containing in total 5404 volumes of 3096 patients. In a second step, we distinguish features between the different diseases in the latent space of this network, and use them to classify new data. 

\end{abstract}

\begin{figure}[H]
\begin{framed}
\begin{center}

\includegraphics[scale=0.07]{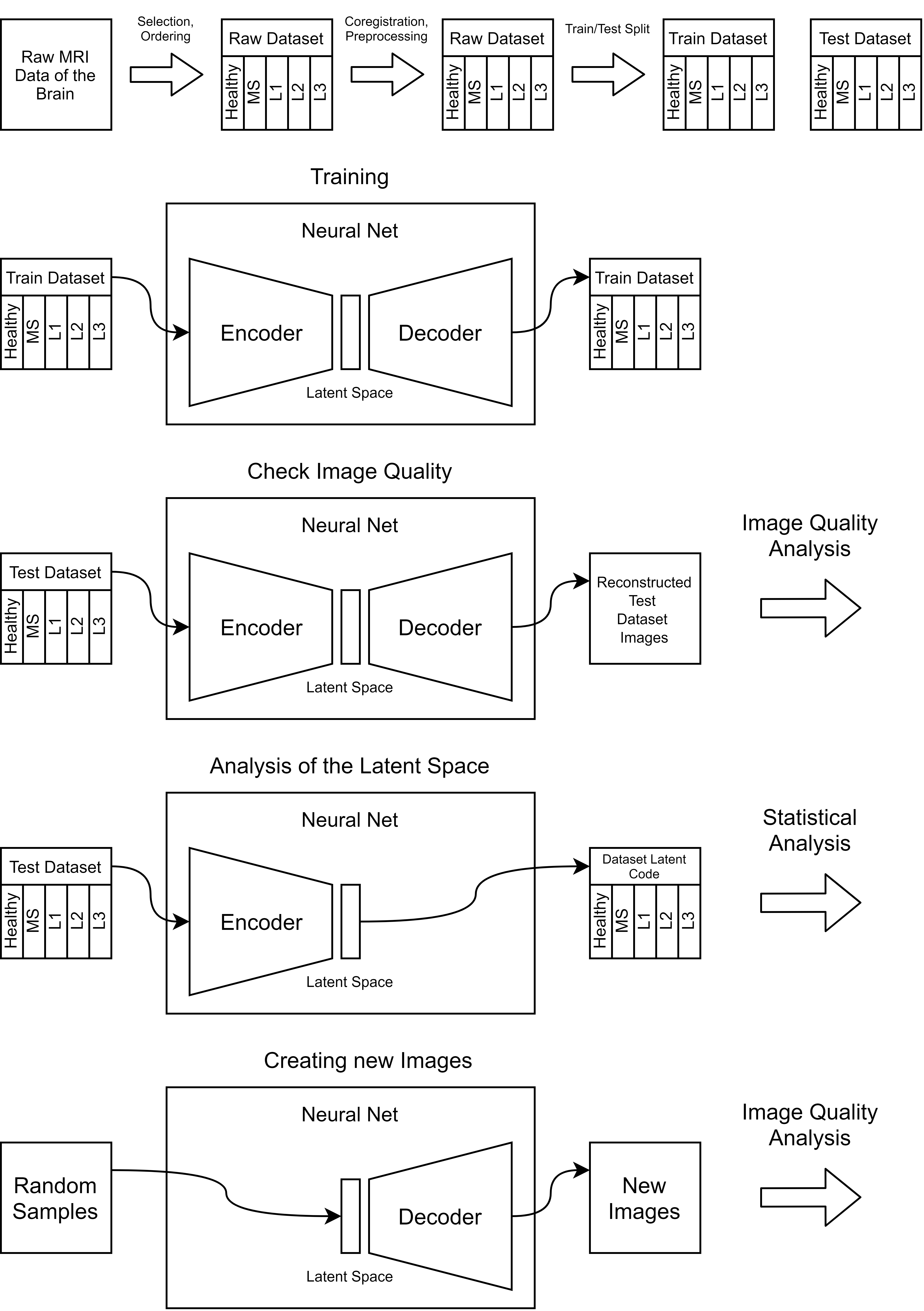}
 \caption{Overview}
 \end{center}
 \end{framed}
 \end{figure}

\section{Introduction}
Multiple sclerosis (MS) is a common neurological disease that is characterized by recurring episodes of inflammation in the central nervous system, during which significant demyelination and axonal loss occur. Typical symptoms include muscle weakness and vision, sensation and coordination disorders. MS white matter lesions have a particular pattern on brain magnetic resonance images (MRI), which is used for diagnosis and follow-up of the disease.\cite{radiologyassistant:MS} Microvascular leukoencephalopathy is a collective term for chronic ischemic white matter brain lesions due to microvascular disease in the context of atherosclerosis.\cite{Leukodystrophies} Both conditions lead to signal anomalies on Fluid Attenuated Inversion Recovery (FLAIR) magnetic resonance (MR) images, which can be distinguished by an expert neuroradiologist, by differentiating subtle differences in the number, aspect, anatomic location and distributions of the lesions \cite{Thompson_diagnosticcriteria}, but not using simple criteria such as for example a signal threshold. The lesions can look very similar to the untrained eye as well as in the early stage of both diseases. 
\\
We used generative network models to reconstruct and generate FLAIR MRI data. Specifically, we trained Variational Autoencoders (VAE) and an Introspective Autoencoders (Intro-VAE) on a mixed database of normal, multiple sclerosis and leukoencephalopathy MRI scans and compared the generated images in terms of their quality. The networks took whole volumes as three dimensional input. We then decomposed and analyzed the latent space of the networks using LDA in order to check if typical lesions characteristics of the MS pattern can be found encoded in the latent space and if we can use the latent space to distinguish between images of the three conditions.

\section{Related Work}
Neural network development in computer vision happened mostly on non-volumetric 2D images, and similarly, most applications of neural networks to medical image analysis was done in two dimensions. There are a multitude of papers applying neural network models on MS datasets, and the developed networks are usually not meant to distinguish between lesions caused by different disease.
\cite{1811.02942} \cite{1803.09172} \cite{1901.05733} \cite{1904.03041} 
\\\\
Generative models have also been sparsely applied to medical imaging of MS. Looking at disease detection, in one paper, the authors trained a VAE only on the data of healthy patients, with the idea that abnormal samples, in their case images with MS lesions, would also have an abnormal latent space code\cite{1806.02997}. They found their approach successful, but this approach is not disease specific, and a brain image with a stroke lesion or other irregularity would also be detected as abnormal. 
\\\\
Three dimensional brain images, from healthy data but also with tumors and strokes, could be generated with different GAN architectures. \cite{1908.02498} A different paper employed a Vector-Quantised Variational Autoencoders (VQ-VAE) on 3 dimensional MRI data. \cite{tudosiu2020neuromorphologicalypreserving} They achieved impressive results in image compression and reconstruction, but didn't do an analysis of the latent space nor did they show newly sampled brains only reconstructed ones.
\\\\

 \section{Theory}
Generative models are machine learning models that try to estimate the joint distribution $P_{Y,X} (y,x)$ of the data $X$ and the target variable $Y$. This allows them to generate new data which is coherent with the dataset. There exist multiple neural network architectures that use generative approaches. Two common ones are Variational Autoencoders and Generative Adversarial Networks.\cite{1701.00160} 

\subsection{Variational Autoencoder (VAE)}
Variational autoencoders (VAEs) consist of an encoding part and a decoding part, which are trained jointly. VAEs share some similarities with basic autoencoders \cite{Goodfellow-et-al-2016}  but are a type of generative model and use a vector of latent random variables. Typically, their target distribution is a (multivariate) gaussian: the encoder compresses the high dimensional input into a mean vector $\mu$ and standard deviation $\sigma$ vector with lower dimensions than the input. A sample $z$, also called a latent vector, can then be fed to the decoder to restore the data as good as possible. To generate new data one can simply sample new latent vectors from the target distribution and pass them through the decoder network.
\\
The objective function of the VAE consists of two terms.\cite{1312.6114} The first term tries to minimize the reconstruction error. The second term is a regularizer that tries to match the distribution of the latent variables generated by encoding the data with a chosen distribution over the latent variables:
$$
L(\phi,\theta;x) = \mathbb{E}_{q_{\phi}(z|x)} [\log( p_{\phi} (x|z))] - KL[q_{\phi}(z|x)||p_{\phi}(z)]
$$
where $x$: datapoint, $z$: latent variable, $p_{\phi} (z)$: true distribution of the latent variable, 
$q_{\phi} (z|x)$: simple distribution of the latent variable given the data (typically $q_{\phi} (z|x)=\mathcal{N}(z;\mu,\sigma^2 I))$\\

\subsection{Generative Adversarial Network (GAN)}
Generative Adversarial Networks (GANs), another popular form of generative models, are composed of two networks that are trained jointly: A generator that uses random noise to create new datapoints and a discriminator that tries to tell the generated and real data apart.\cite{1406.2661}. In contrast to VAEs, GANs try to model the data distribution directly instead of approximating it with a chosen target distribution. 
The GAN tries to solve the following optimization problem: \cite{gan:eq}
$$
\min_{G}\max_{D}\mathbb{E}_{x \thicksim p_{data}(x)}[\log (D(x))] + \mathbb{E}_{z \thicksim p_{z}(z)}[log(1-D(G(z)))]
$$
where $D(x)$: discriminator's estimate of the probability that real datapoint $x$ is real, $G(z)$: generator's output given noise $z$, $D(G(z))$: discriminator's estimate of the probability that a fake instance is real
\\\\
To create new data with a GAN, one simply feeds the generator a random vector of noise. Typically GANs produce sharper images compared to VAEs. Indeed, a lack of sharpness in the generated images would be used by the discriminator to identify the generated images.

\subsection{Introspective Variational Autoencoder} 
The introspective variational autoencoder (intro-VAE or IVAE) is a modification of a standard VAE. With the idea to combine the strength of VAEs, namely their nice manifold representations in the latent space, with the strength of GANs, their sharpness of generated images. But compared to a GAN the sampling diversity should be improved and the training should be more stable.\cite{1807.06358}
\\
The IVAE has the same internal construction as a VAE, but its encoder simultaneously acts as discriminator and its decoder can also be thought of as generator. This is the main distinguishing factor from other VAE and GAN combinations which typically rely on a separate discriminator.
\\
The paper introducing the IVAE states that the blurriness of VAEs originates from the "assignment of high probability to training points"\cite{1807.06358} and claims a VAE "cannot ensure that a low probability is assigned to blurry datapoints"\cite{1807.06358}. The IVAE has the possibility to change the probability given to blurry points since it can behave like a normal VAE for real data but acts as a GAN in the case of generated data. \cite{1807.06358}. 
\\
The training is similar to that of the GAN but the loss for the encoder and generator contain also the VAE loss.  
\\
$$
L_{E}(x,z) =  E(x) + max(0, m - E(G(z))) + L_{AE}(x)
$$
$$
L_{G}(z) = E(G(z)) + L_{AE}(x)
$$
where $E(x) = KL(q_{\Phi}(z|x)||p(z))$, with $x$ being a datapoint, $G(z)$: generator's output given noise $z$, $L_{AE}(x) = \mathbb{E}_{q_{\phi}(z|x)} [\log p_{\phi} (x|z)]$

\subsection{Linear Discriminant Analysis}
The Linear Discriminant Analysis (LDA) is a method to find a subspace of a feature space that maximizes the separability between classes. As such the LDA can be used to show separability of high dimensional data and to predict the class of a datapoint. \cite{sebastianraschka:LDA}

\section{Methods}

\subsection{Databases}

Three databases were produced: a normal database, a multiple sclerosis database and a leukoencephalopathy database which differentiated between three degrees of severity.

\subsubsection{Databases Characteristics}
All three databases consisted exclusively of MR FLAIR scans.
Together the databases contained 5404 MR scans of 3096 patients.
No Patient was in multiple databases.
\\
1855 scans from 1855 patients were included in the normal database; patient age: [mean $\pm$ standard deviation] 39 $\pm$ 24 y.
2910 scans from 616 patients were included in the MS database; patient age: 46 $\pm$ 14 y.
639 scans from 625 patients were included in the leukoencephalopathy database:
(393 scans from 384 patients were of severity 1, 41 scans from 40 patients were of severity 2, 205 scans from 201 patients were of severity 3); patient age: 75 $\pm$ 10 y.
Figure \ref{fig:classes} shows an overview over the databases.

\begin{table}[h]
\begin{framed}
\begin{tabularx}{1\textwidth} { 
      | >{\raggedright\arraybackslash}X 
      | >{\raggedleft\arraybackslash}X 
      | >{\raggedleft\arraybackslash}X 
      | >{\raggedleft\arraybackslash}X | }
     \hline
      & Number of Patients & Number of MR images & Age [mean $\pm$ standard deviation]\\
     \hline
     Healthy  & 1855  & 1855 & 39 $\pm$ 24\\
     \hline
     MS & 616 & 2910 & 46 $\pm$ 14\\
     \hline
     L 1 & 384 & 393 & 73 $\pm$ 10\\
     \hline
     L 2 & 40 & 41 & 76 $\pm$ 9\\
     \hline
     L 3 & 201 & 205 & 81 $\pm$ 8\\
     \hline
\end{tabularx}
\caption{Data distribution over classes. L is short for leukoencephalopathy}
\label{fig:classes}
\end{framed}
\end{table}

\subsection{Data pre-processing}
\subsubsection{Coregistration}
In a first step the MR scans were coregistered.
The framework used for the coregistration was SimpleElastix \cite{SimpleElastix:tool}, a python wrapper for the Elastix framework \cite{elastix:tool}  which uses the itk framework \cite{ITK:tool}. For skull stripping the fsl BET \cite{fsl:tool} tool was used.
\\
The coregistration parameters were:
\begin{itemize}
\item Skull-stripped template and skull-stripped moving image
\item Affine transformation to the MNI template \cite{template}
\item Advanced Mattes Mutual Information
\item Output dimensions: 182$\times$218$\times$182
\end{itemize}
\begin{figure}[H]
\begin{framed}
\includegraphics[height=120 px]{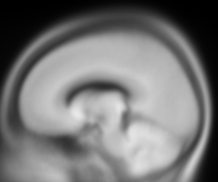}
\includegraphics[height=120 px]{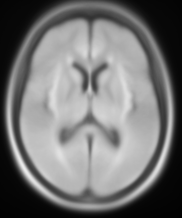}
\includegraphics[height=120 px]{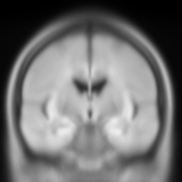}
\caption{Mean Image over the whole dataset, coregistered with Advanced Mattes Mutual Information, skull stripped template and skull stripped moving image}
\label{fig:images/CoReg/coreg2.png}
\end{framed}
\end{figure}
\noindent
We checked the result of the coregistration with a mean image (Figure \ref{fig:images/CoReg/coreg2.png}) which was computed over all coregistered images in the dataset. The sharpness of this mean image was used as quality measure.

\subsubsection{Trimming and down sampling}
In a second step the MR scans were trimmed to a resolution of 160$\times$192$\times$160 and downsampled to the final resultion of 40$\times$48$\times$40 by taking the average over 4$\times$4$\times$4 voxels.

\subsubsection{Bounding the voxel values}
In a last step the upper value of voxels in an image was constrained to the value of the 99.5 percentile and the voxel values were transformed into the interval [0, 1] using the following formula.
$$
X_{new} = \frac{X_{old}-\min(X)}{\max(X)}
$$
\centerline{$X$: voxel array on an image}

\subsection{Datasets for Training and Testing}
All three databases were randomly divided into a training set (90\%) and a test set (10\%). 
To prevent cross contamination this was done on the level of patients and not on the level of images. 
The final training dataset contained the training sets of all three databases.
All neural network models were trained on the same final training set and analysed using the same final test set.

\subsection{Neural Networks}
We created four different neural networks using Keras with a TensorFlow backend. Two were VAE models which had latent space dimensionality 8 and 512 referred to as VAE-8 and VAE-512. Two were IVAE models which had latent space dimensionality 32 and 256 referred to as IVAE-32 and IVAE-256 respectively.
\\
All four neural network models produced shared the same overall structure for the encoder and decoder. They only differed in the size of the latent space.
\\
The networks input and output were whole three-dimensional MR images, not just two-dimensional slices. The exact input and output dimensions were 40$\times$48$\times$40. The encoder used multiple convolutional layers with batch normalization and added pooling to reduce the resolution down to 5$\times$6$\times$5$\times$64. Dense layers with dropout and batch normalization reduced the number of parameters further to the latent dimension.
\\
The decoder we used was a structural reversal of the encoder. It consisted of dense layers with dropout and batch normalization that increased the number of parameters from the latent dimension up to 5$\times$6$\times$5$\times$64 again. After that deconvolutional layers with batch normalization and added upsampling were utilized to reach the output resolution.
\\
A NVIDIA Titan RTX with 24 GB of memory was used to train the models. The models were trained for one week each.

\section{Results}

Four neural network models (VAE-8, VAE-512, IVAE-32 and IVAE-256) were trained on the same dataset.
We evaluated the performance of the networks in terms of image reconstruction and generation. 
The created images were compared to images from the database to determine which features the neural networks were able to recreate and which were missing. 
We analysed the separability of image labels based on the associated latent space.

\subsection{Reconstruction quality}
\begin{figure}[H]
\begin{framed}
\begin{center}
\begin{tabular}{ c | c | c | c | c }

 \includegraphics[scale=0.38]{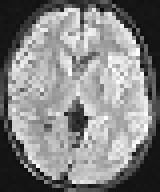} & 
 \includegraphics[scale=0.38]{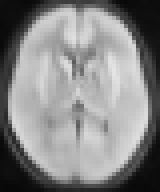} & 
 \includegraphics[scale=0.38]{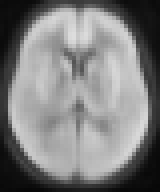} & 
 \includegraphics[scale=0.38]{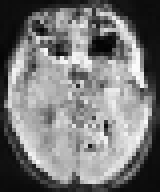} & 
 \includegraphics[scale=0.38]{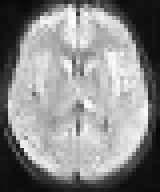} \\ 
 
 \includegraphics[scale=0.38]{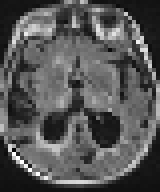} & 
 \includegraphics[scale=0.38]{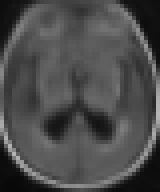} & 
 \includegraphics[scale=0.38]{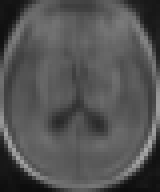} & 
 \includegraphics[scale=0.38]{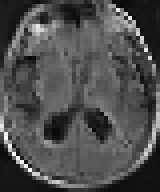} & 
 \includegraphics[scale=0.38]{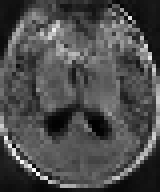} \\ 
 
 \includegraphics[scale=0.38]{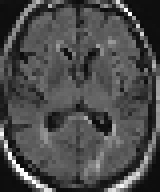} & 
 \includegraphics[scale=0.38]{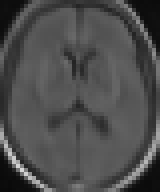} & 
 \includegraphics[scale=0.38]{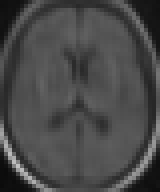} & 
 \includegraphics[scale=0.38]{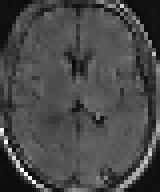} & 
 \includegraphics[scale=0.38]{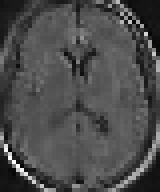} \\ 
 
 \includegraphics[scale=0.38]{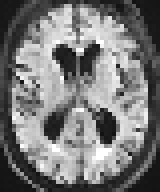} & 
 \includegraphics[scale=0.38]{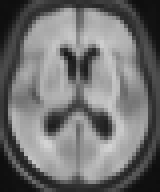} & 
 \includegraphics[scale=0.38]{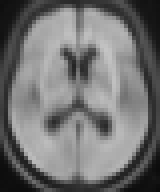} & 
 \includegraphics[scale=0.38]{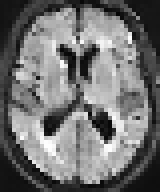} & 
 \includegraphics[scale=0.38]{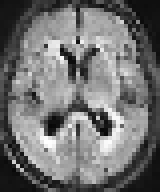} \\ 
 
 \includegraphics[scale=0.38]{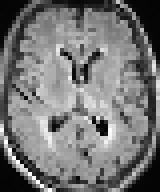} & 
 \includegraphics[scale=0.38]{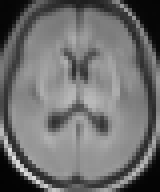} & 
 \includegraphics[scale=0.38]{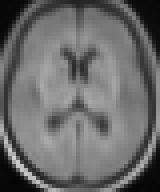} & 
 \includegraphics[scale=0.38]{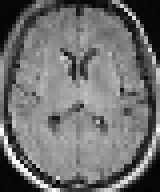} & 
 \includegraphics[scale=0.38]{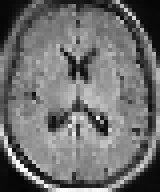} \\ 
 
 \includegraphics[scale=0.38]{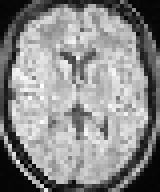} & 
 \includegraphics[scale=0.38]{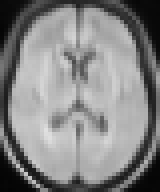} & 
 \includegraphics[scale=0.38]{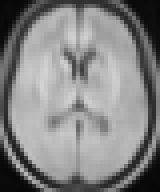} & 
 \includegraphics[scale=0.38]{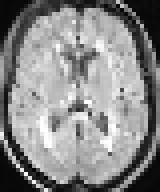} & 
 \includegraphics[scale=0.38]{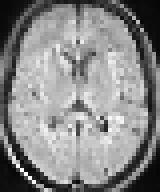} \\ 
 
 Original & VAE-8 & VAE-512 & IVAE-32 & IVAE-256\\
 \hline
 
\end{tabular}
\end{center}
\caption{Axial MRI, on the left the original image, in the other columns the reconstruction of the mentioned model}
\label{tab:Axial_MRI}
\end{framed}
\end{figure}

\begin{figure}[H]
\begin{framed}
\begin{center}
\begin{tabular}{ c | c | c | c | c }

 \includegraphics[scale=0.38]{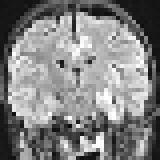} & 
 \includegraphics[scale=0.38]{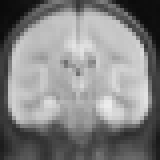} & 
 \includegraphics[scale=0.38]{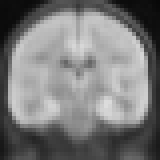} & 
 \includegraphics[scale=0.38]{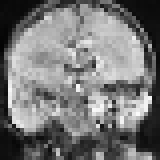} & 
 \includegraphics[scale=0.38]{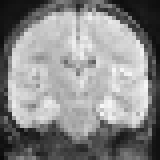} \\ 
 
 \includegraphics[scale=0.38]{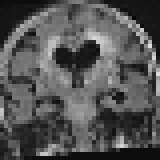} & 
 \includegraphics[scale=0.38]{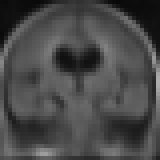} & 
 \includegraphics[scale=0.38]{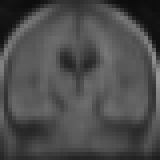} & 
 \includegraphics[scale=0.38]{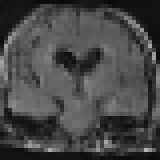} & 
 \includegraphics[scale=0.38]{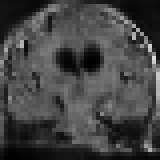} \\ 
 
 \includegraphics[scale=0.38]{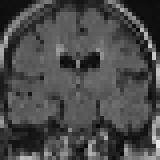} & 
 \includegraphics[scale=0.38]{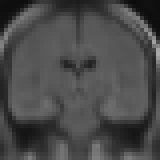} & 
 \includegraphics[scale=0.38]{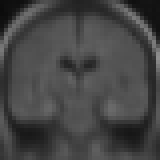} & 
 \includegraphics[scale=0.38]{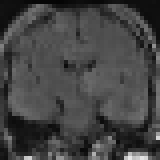} & 
 \includegraphics[scale=0.38]{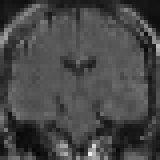} \\ 
 
 \includegraphics[scale=0.38]{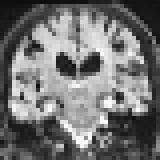} & 
 \includegraphics[scale=0.38]{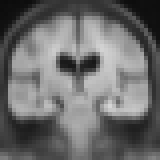} & 
 \includegraphics[scale=0.38]{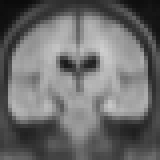} & 
 \includegraphics[scale=0.38]{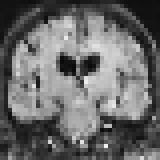} & 
 \includegraphics[scale=0.38]{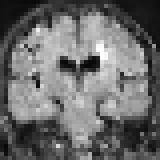} \\ 
 
 \includegraphics[scale=0.38]{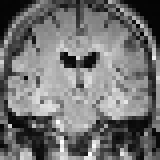} & 
 \includegraphics[scale=0.38]{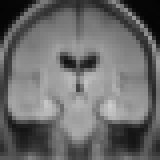} & 
 \includegraphics[scale=0.38]{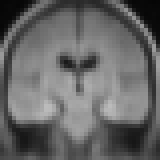} & 
 \includegraphics[scale=0.38]{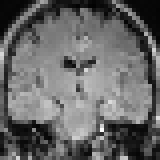} & 
 \includegraphics[scale=0.38]{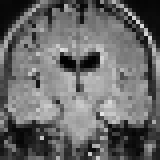} \\ 
 
 \includegraphics[scale=0.38]{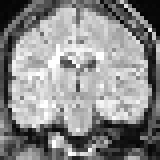} & 
 \includegraphics[scale=0.38]{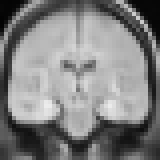} & 
 \includegraphics[scale=0.38]{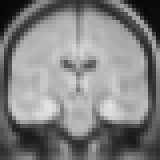} & 
 \includegraphics[scale=0.38]{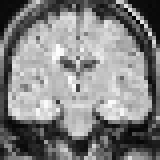} & 
 \includegraphics[scale=0.38]{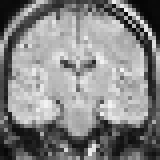} \\ 
 
 Original & VAE-8 & VAE-512 & IVAE-32 & IVAE-256 \\
 \hline
 
\end{tabular}
\end{center}
\caption{Coronal MRI, on the left the original image, in the other columns the reconstruction of the mentioned model}
\label{tab:Coronal_MRI}
\end{framed}
\end{figure}

\begin{figure}[H]
\begin{framed}
\begin{center}
\begin{tabular}{ c | c | c | c | c }

 \includegraphics[scale=0.3]{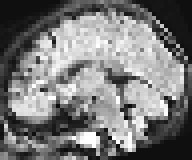} & 
 \includegraphics[scale=0.3]{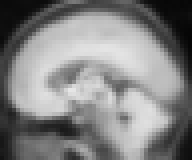} & 
 \includegraphics[scale=0.3]{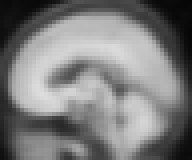} & 
 \includegraphics[scale=0.3]{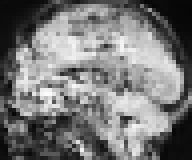} & 
 \includegraphics[scale=0.3]{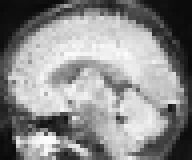} \\ 
 
 \includegraphics[scale=0.3]{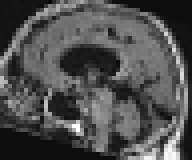} & 
 \includegraphics[scale=0.3]{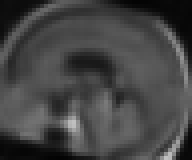} & 
 \includegraphics[scale=0.3]{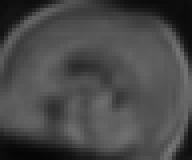} & 
 \includegraphics[scale=0.3]{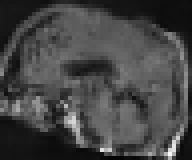} & 
 \includegraphics[scale=0.3]{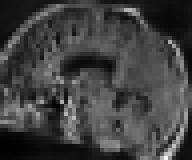} \\ 
 
 \includegraphics[scale=0.3]{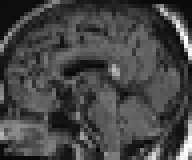} & 
 \includegraphics[scale=0.3]{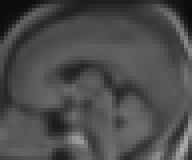} & 
 \includegraphics[scale=0.3]{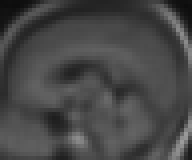} & 
 \includegraphics[scale=0.3]{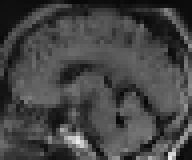} & 
 \includegraphics[scale=0.3]{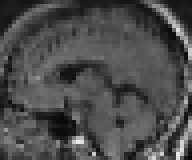} \\ 
 
 \includegraphics[scale=0.3]{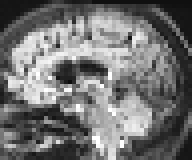} & 
 \includegraphics[scale=0.3]{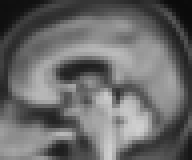} & 
 \includegraphics[scale=0.3]{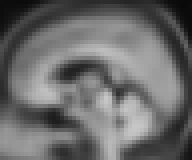} & 
 \includegraphics[scale=0.3]{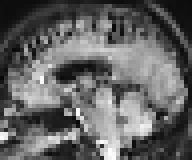} & 
 \includegraphics[scale=0.3]{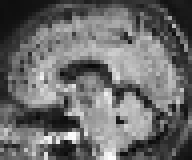} \\ 
 
 \includegraphics[scale=0.3]{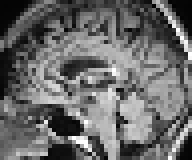} & 
 \includegraphics[scale=0.3]{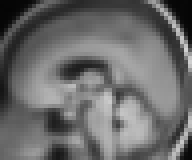} & 
 \includegraphics[scale=0.3]{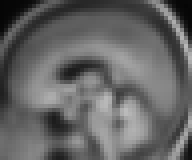} & 
 \includegraphics[scale=0.3]{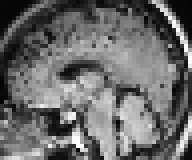} & 
 \includegraphics[scale=0.3]{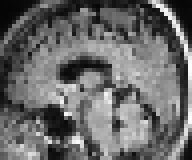} \\ 
 
 \includegraphics[scale=0.3]{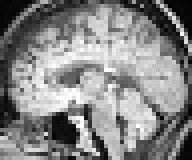} & 
 \includegraphics[scale=0.3]{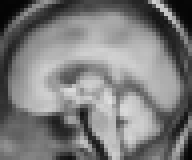} & 
 \includegraphics[scale=0.3]{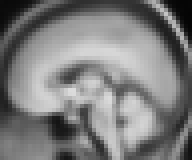} & 
 \includegraphics[scale=0.3]{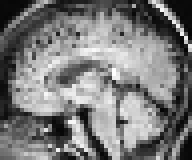} & 
 \includegraphics[scale=0.3]{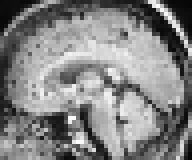} \\ 
 
 Original & VAE-8 & VAE-512 & IVAE-32 & IVAE-256 \\
 \hline
 
\end{tabular}
\end{center}
\caption{Sagittal MRI, on the left the original image, in the other columns the reconstruction of the mentioned model}
\label{tab:Sagittal_MRI}
\end{framed}
\end{figure}

The figures \ref{tab:Axial_MRI}, \ref{tab:Coronal_MRI} and \ref{tab:Sagittal_MRI} compare the image reconstruction quality of the different models. The leftmost column shows a slice of the original image, which is from the test set. The other four columns show the reconstruction of that image from the respective model.
\\\\
Both VAE models produced very blurry results. Only the lateral ventricle, the dark colored part in the middle of the brain, and the overall color scheme was reconstructed with mentionable accuracy. The difference of white and grey matter was only hinted at and certainly not accurate. The VAE models also did not show any gyri, the skin folds on the surface of the brain, in the reconstruction.
\\\\
The IVAE models created sharper reconstructions but they had inaccuracies. The lateral ventricle was in some examples poorly reconstructed. In some places there were some holes or dark spots that weren’t there in the original image. The difference of white and grey matter was hinted at but was still rather blurry. Same was true for the gyri that showed up in the reconstructions as a bunch of darker pixels on the edges of the brain but lack structure.
\\\\
For both model types the differences in terms of image quality between the model with few and the model with many latent space dimensions were sometimes noticeable but often minimal.

\subsection{Image Generation Quality}
\begin{figure}[H]
\begin{framed}
\begin{center}
\begin{tabular}{ c | c | c | c | c  }

 \includegraphics[width=60px]{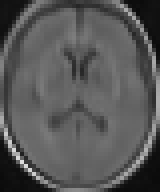} & 
 \includegraphics[width=60px]{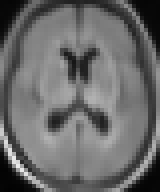} & 
 \includegraphics[width=60px]{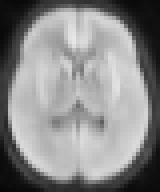} & 
 \includegraphics[width=60px]{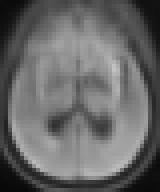} & 
 \includegraphics[width=60px]{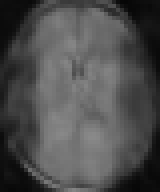}
 \\

 \includegraphics[width=60px]{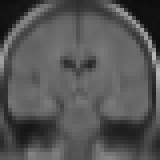} & 
 \includegraphics[width=60px]{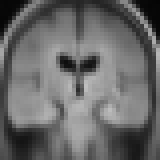} & 
 \includegraphics[width=60px]{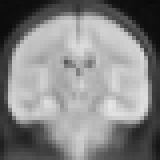} & 
 \includegraphics[width=60px]{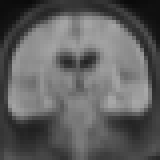} & 
 \includegraphics[width=60px]{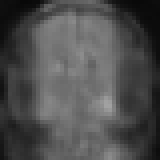}\\

 \includegraphics[width=60px]{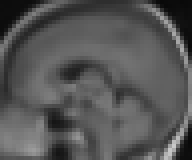} & 
 \includegraphics[width=60px]{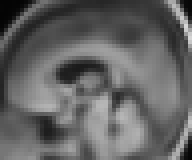} & 
 \includegraphics[width=60px]{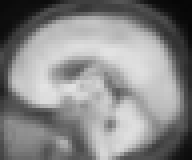} & 
 \includegraphics[width=60px]{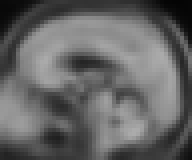} & 
 \includegraphics[width=60px]{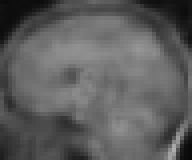}\\

 VAE-8 & VAE-8 & VAE-8 & VAE-8 & VAE-8\\
 \hline
 
\end{tabular}
\end{center}
\caption{axial, coronal and sagittal MRI slice, sampled by the VAE network with latent space dimension 8.}
\label{tab:gen_VAE-8}
\end{framed}
\end{figure}

\begin{figure}[H]
\begin{framed}
\begin{center}
\begin{tabular}{ c | c | c | c | c  }

 \includegraphics[width=60px]{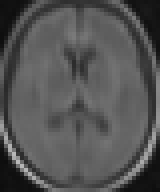} & 
 \includegraphics[width=60px]{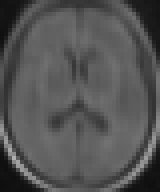} & 
 \includegraphics[width=60px]{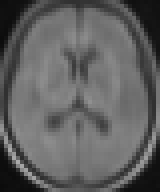} & 
 \includegraphics[width=60px]{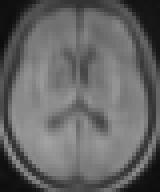} & 
 \includegraphics[width=60px]{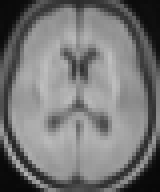}
 \\

 \includegraphics[width=60px]{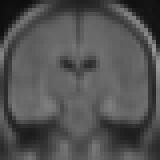} & 
 \includegraphics[width=60px]{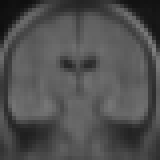} & 
 \includegraphics[width=60px]{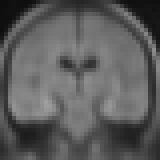} & 
 \includegraphics[width=60px]{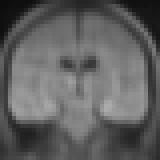} & 
 \includegraphics[width=60px]{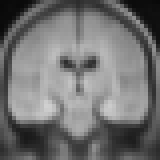}
 \\

 \includegraphics[width=60px]{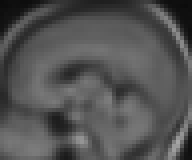} & 
 \includegraphics[width=60px]{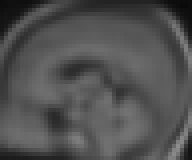} & 
 \includegraphics[width=60px]{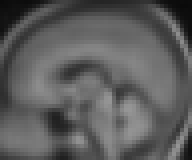} & 
 \includegraphics[width=60px]{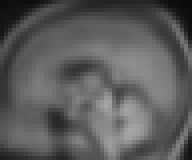} & 
 \includegraphics[width=60px]{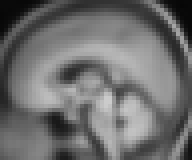}
 \\

 VAE-512 & VAE-512 & VAE-512 & VAE-512 & VAE-512\\
 \hline
 
\end{tabular}
\end{center}
\caption{axial, coronal and sagittal MRI slice, sampled by the VAE network with latent space dimension 512.}
\label{tab:gen_VAE-512}
\end{framed}
\end{figure}

\begin{figure}[H]
\begin{framed}
\begin{center}
\begin{tabular}{ c | c | c | c | c  }

 \includegraphics[width=60px]{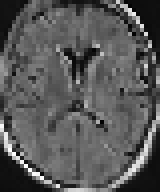} & 
 \includegraphics[width=60px]{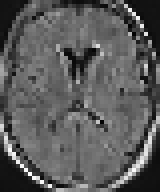} & 
 \includegraphics[width=60px]{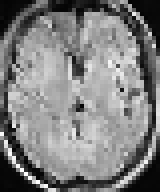} & 
 \includegraphics[width=60px]{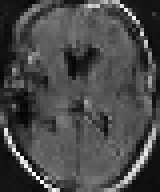} & 
 \includegraphics[width=60px]{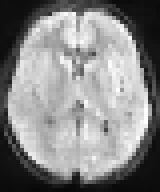}
 \\

  \includegraphics[width=60px]{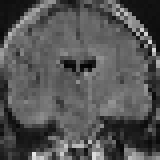} & 
 \includegraphics[width=60px]{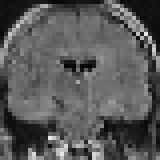} & 
 \includegraphics[width=60px]{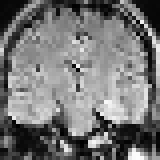} & 
 \includegraphics[width=60px]{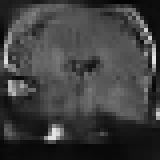} & 
 \includegraphics[width=60px]{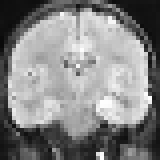}
 \\

  \includegraphics[width=60px]{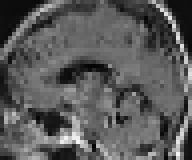} & 
 \includegraphics[width=60px]{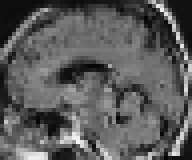} & 
 \includegraphics[width=60px]{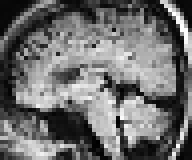} & 
 \includegraphics[width=60px]{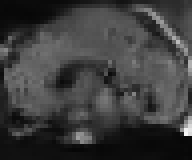} & 
 \includegraphics[width=60px]{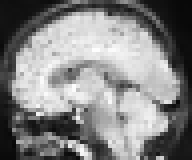}
 \\

 IVAE-32 & IVAE-32 & IVAE-32 & IVAE-32 & IVAE-32\\
 \hline
 
\end{tabular}
\end{center}
\caption{axial, coronal and sagittal MRI slice, sampled by the IVAE network with latent space dimension 32.}
\label{tab:gen_IVAE-32}
\end{framed}
\end{figure}

\begin{figure}[H]
\begin{framed}
\begin{center}
\begin{tabular}{ c | c | c | c | c  }

 \includegraphics[width=60px]{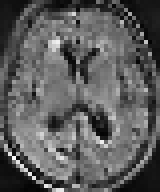} & 
 \includegraphics[width=60px]{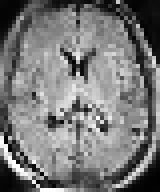} & 
 \includegraphics[width=60px]{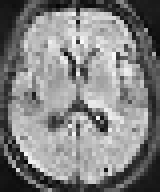} & 
 \includegraphics[width=60px]{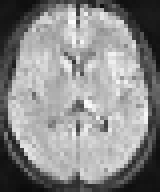} & 
 \includegraphics[width=60px]{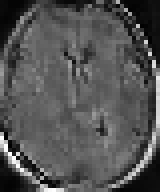}
 \\

  \includegraphics[width=60px]{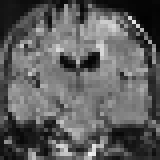} & 
 \includegraphics[width=60px]{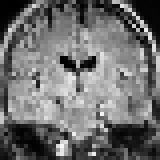} & 
 \includegraphics[width=60px]{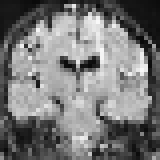} & 
 \includegraphics[width=60px]{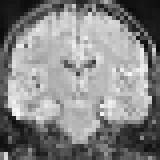} & 
 \includegraphics[width=60px]{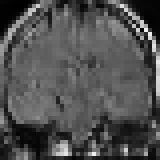}
 \\

  \includegraphics[width=60px]{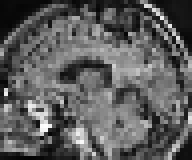} & 
 \includegraphics[width=60px]{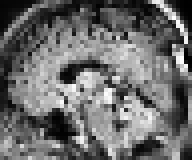} & 
 \includegraphics[width=60px]{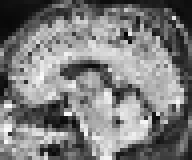} & 
 \includegraphics[width=60px]{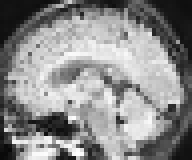} & 
 \includegraphics[width=60px]{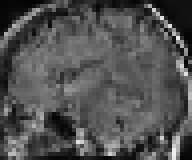}
 \\

 IVAE-256 & IVAE-256 & IVAE-256 & IVAE-256 & IVAE-256\\
 \hline
 
\end{tabular}
\end{center}
\caption{axial, coronal and sagittal MRI slice, sampled by the IVAE network with latent space dimension 256.}
\label{tab:gen_IVAE-256}
\end{framed}
\end{figure}

The figures \ref{tab:gen_VAE-8}, \ref{tab:gen_VAE-512}, \ref{tab:gen_IVAE-32} and \ref{tab:gen_IVAE-256} compare the image generation quality. The axial, coronal and sagittal slice that are grouped together belong to the same brain. The latent space vectors were drawn from a standard normal distribution which was the target distribution of the models during training.\\\\
As in the image reconstruction part, the IVAE models once again returned sharper images than the VAE models.\\\\ 
For the VAE models aside from the changing color scheme and some small changes to the lateral ventricle most of the image stayed the same over different latent samples.\\\\
The IVAE models produced a bigger variety of different brains although their quality wasn't on par with the reconstructed ones. In some examples they created unnatural looking artifacts. The lateral ventricles were vanishing in some images and looked worse than in the reconstructed examples. Some images also seemed patchy and incoherent.\\\\
In most cases neither the VAE nor the IVAE models created sharp images of new brains.

\subsection{Latent space}
\begin{figure}[H]
\begin{framed}
\begin{center}
\begin{tabular}{ c }

\includegraphics[scale=0.267]{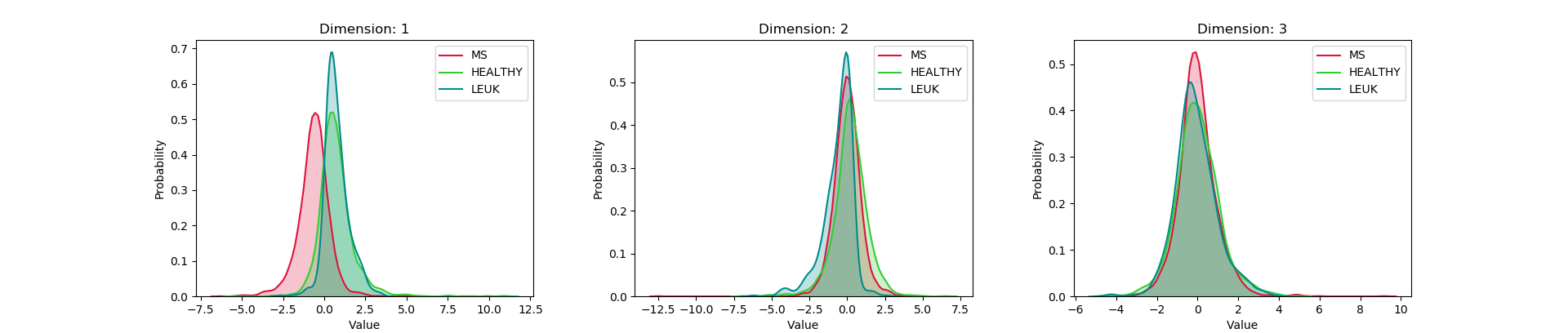}
\\
VAE model with latent dimenson 8
\\
\includegraphics[scale=0.267]{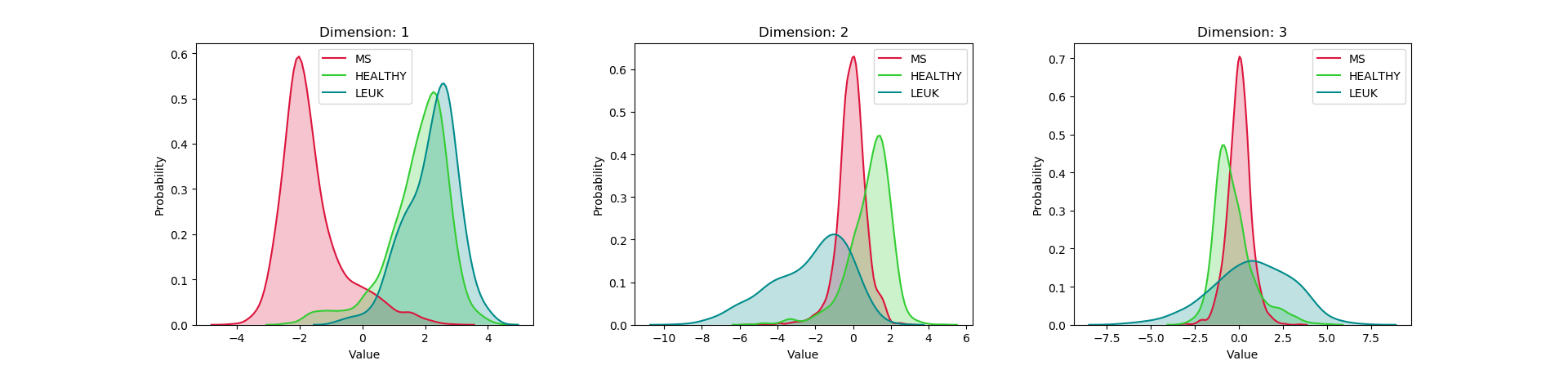}
\\
VAE model with latent dimenson 512
\\
\includegraphics[scale=0.267]{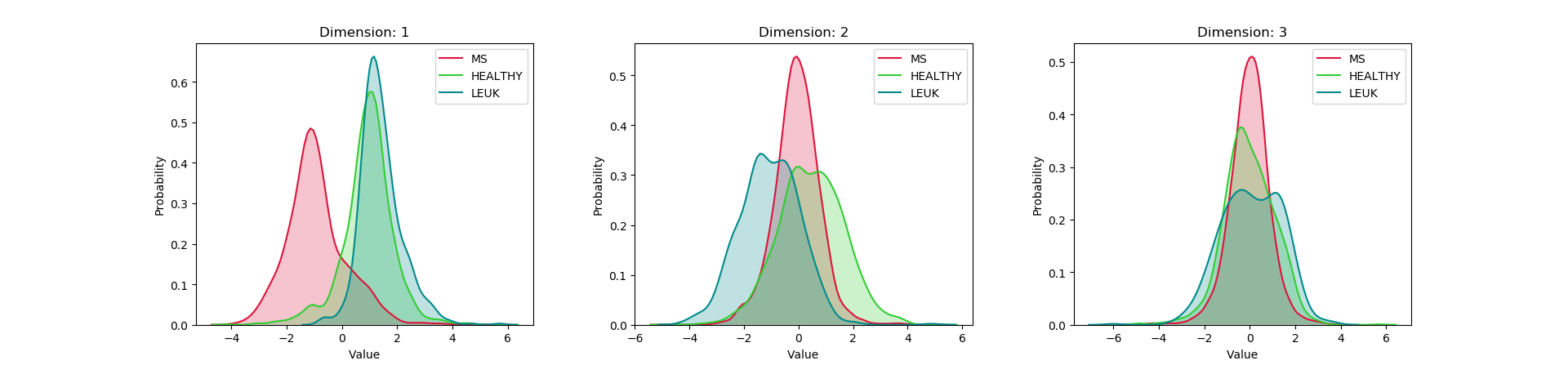}
\\
IVAE model with latent dimenson 32
\\
\includegraphics[scale=0.267]{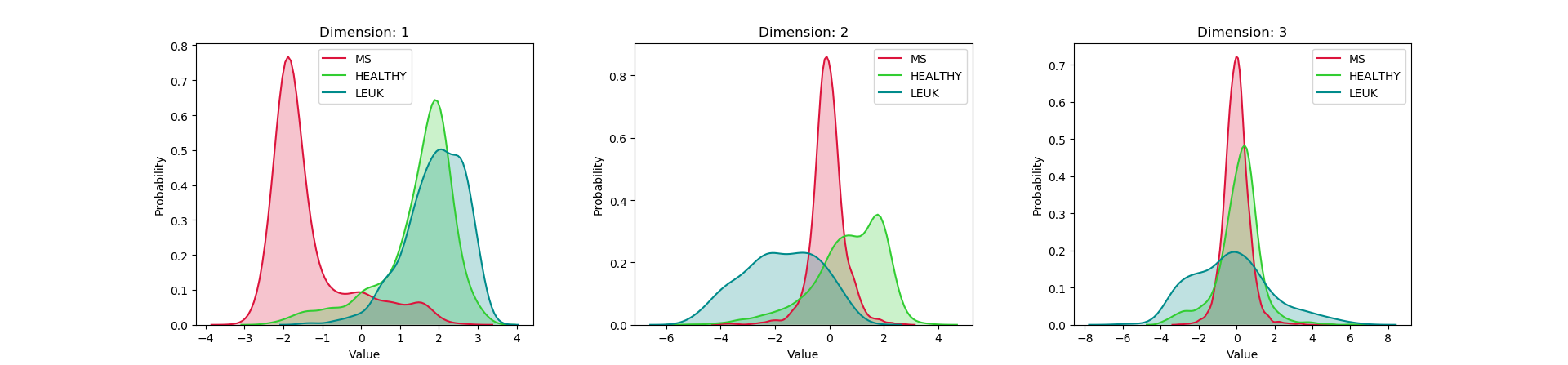}
\\
IVAE model with latent dimenson 256
\\
\end{tabular}
\end{center}
\caption{The first three dimensions of the LDA composition done for the different models on latent space vectors of the test set. The models with a larger latent space show a good separability for the MS images from images of the other two classes. While the separability between healthy and leukoencephalopathy images is not so clean.}
\label{tab:LDA}
\end{framed}
\end{figure}

Figure \ref{tab:LDA} shows a linear discriminant analysis (LDA) of the latent space of the different models. To improve readability of the plots the three leukoencephalopathy groups were accumulated into a single category. We show the first three dimensions of the LDA rather than the full latent space since most latent space features had no or very little difference between the distributions of the different classes.
\\
For all four models the LDA has a dimension that showed a good separability between the MS data and the other two categories. The models with a larger latent space also had a dimension that differentiates between healthy and leukoencephalopathy data. The two models with the smaller latent space still showed some separability between healthy and leukoencephalopathy but it was not as pronounced.

\noindent
Table \ref{table:classification} shows the resulting statistics from an LDA done on the latent space of the IVAE-256 model. The LDA was done on the training set and then used to predict the classes of the test set. The high accuracy and recall for detecting MS stuck out especially but the accuracy and recall for the healthy data was still good. Only the three leukoencephalopathy classes were not as often classified correctly.
\\\\
\begin{table}[H]
\begin{framed}
\begin{tabularx}{1\textwidth} { 
      | >{\raggedright\arraybackslash}X 
      | >{\raggedleft\arraybackslash}X 
      | >{\raggedleft\arraybackslash}X 
      | >{\raggedleft\arraybackslash}X 
      | >{\raggedleft\arraybackslash}X 
      | >{\raggedleft\arraybackslash}X | }
     \hline
      & MS & Leuk 1 & Leuk 2 & Leuk 3 & Healthy \\
     \hline
     FP  & 24  & 29 & 3 & 13 & 45  \\
     \hline
     FN  & 35  & 18 & 3 & 10 & 48  \\
     \hline
     TP  & 285  & 22 & 1 & 12 & 138  \\
     \hline
     TN  & 228  & 503 & 565 & 537 & 341  \\
     \hline
\end{tabularx}
\\\\ \\\\ 
\begin{tabularx}{1\textwidth} { 
      | >{\raggedright\arraybackslash}X 
      | >{\raggedleft\arraybackslash}X 
      | >{\raggedleft\arraybackslash}X 
      | >{\raggedleft\arraybackslash}X 
      | >{\raggedleft\arraybackslash}X 
      | >{\raggedleft\arraybackslash}X | }
     \hline
      & MS & Leuk 1 & Leuk 2 & Leuk 3 & Healthy \\
     \hline
     Precision  & 0.92  & 0.43 & 0.25 & 0.48 & 0.75  \\
     \hline
     Recall  & 0.89  & 0.55 & 0.25 & 0.55 & 0.74  \\
     \hline
     
\end{tabularx}
\caption{LDA used on the latent space created by the IVAE model with latent dimenson 256 to classify the data in the test set}
\label{table:classification}
\end{framed}
\end{table}

\begin{figure}[H]
\begin{framed}
\includegraphics[scale=0.257]{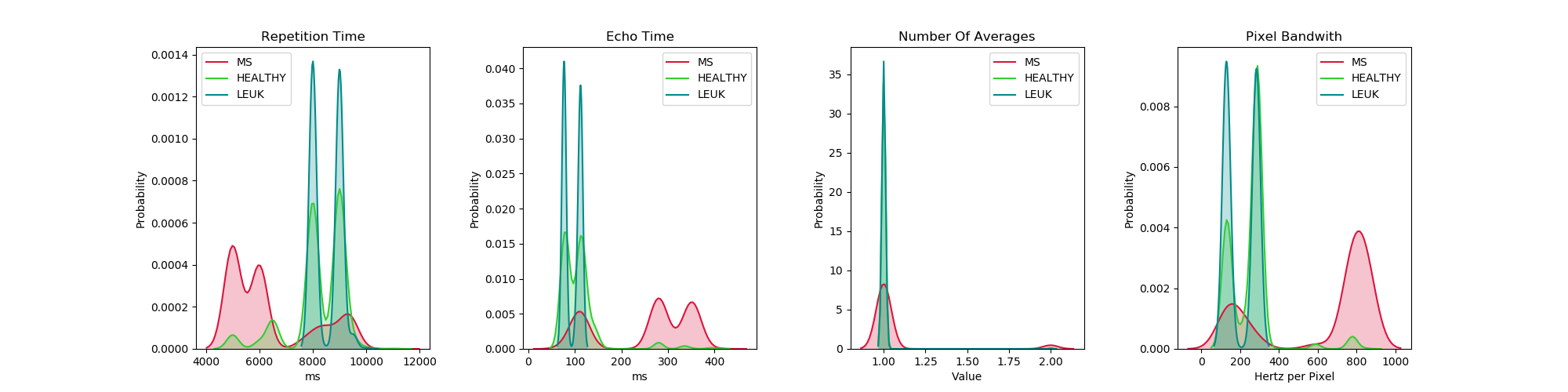}
\caption{Distribution of some metadata attributes of the dataset}
\label{fig:images/graphs/metadata}
\end{framed}
\end{figure}
\noindent
We found that the classes in the dataset correlated with metadata features that can influence the look of the image. Figure \ref{fig:images/graphs/metadata} shows that there is clearly a bias in the dataset. Pixel bandwidth, repetition time and echo time indicate that the MS images were taken with different parameters than images from the other two classes. Both repetition time and echo time are parameters controlling the contrast of the MR image. \cite{mriquestions:contrast} The pixel bandwidth has an influence on the appearance of the image as well. \cite{mrimaster:bandwidth}

\begin{figure}[H]
\begin{framed}
\begingroup
\setlength{\tabcolsep}{2pt}
\begin{tabular}{ c c c c }

 \includegraphics[height=102px]{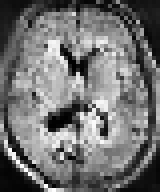}&
\includegraphics[height=102px]{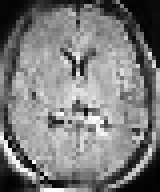}&
\includegraphics[height=102px]{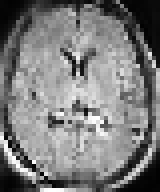}&
\includegraphics[height=102px]{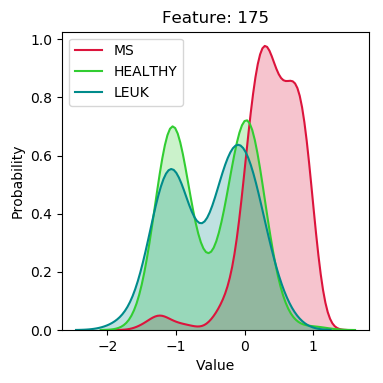}
\\
      Value: -1.25 & Value: 0.00 & Value: 1.25 & Dimension: 175  \\
\hline
\end{tabular}
\endgroup
\caption{Latent Space Feature Dimension on the right, sampled images on the left, taken from the IVAE model with latent dimension 256. All the feature values were kept the same only the value of the 175. dimension was changed.}
\label{fig:images/CoReg/feature_175.png}
\end{framed}
\end{figure}

\begin{figure}[H]
\begin{framed}
\begingroup
\setlength{\tabcolsep}{2pt}
\begin{tabular}{ c c c c }

\includegraphics[height=102px]{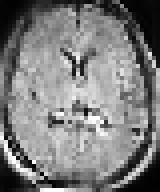}&
\includegraphics[height=102px]{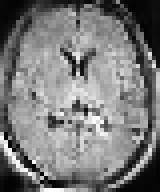}&
\includegraphics[height=102px]{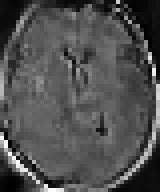}&
\includegraphics[height=102px]{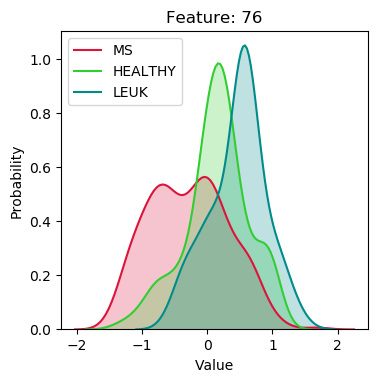}
\\
      Value: -1.00 & Value: 0.25 & Value: 1.00 & Dimension: 76  \\
\hline
\end{tabular}
\endgroup

\caption{Latent Space Feature Dimension on the right, sampled images on the left, taken from the IVAE model with latent dimension 256. All the feature values were kept the same only the value of the 76. dimension was changed.}
\label{fig:images/CoReg/feature_76.png}
\end{framed}
\end{figure}
\noindent
We tried to find out what effect changing values of certain latent space dimensions had on the image. Since the IVAE-256 has a latent space of size 256 we limited ourselves to latent dimension that showed different distributions for the different classes.
\\
Figure \ref{fig:images/CoReg/feature_175.png} and \ref{fig:images/CoReg/feature_76.png} give some insight on what the latent space features change in the image. All the other latent dimensions were set to 0. The first row shows that feature number 175 had an influence on the shape of the lateral ventricle. The second row shows that feature number 76 had an influence on the color scheme of the brain. 
\section{Conclusion}
\subsection{Overview}
We trained and compared VAE and IVAE models with different latent space sizes on a dataset with multiple disease categories. We found that the IVAEs generated more detailed images than the VAEs independent of the latent space size. For the VAEs higher latent space dimensionality seemed to have no beneficial impact on the image quality and the generated images looked always blurry. The IVAEs seemed to profit marginally from having a higher latent space resulting in less artefacts in the generated images.\\
Over all models there was a noticeable quality difference between reconstructed and sampled images, for the VAEs the sampled images were less diverse than the reconstructed ones and for IVAEs there was a noticeable loss of detail for the sampled images.
\\
Examining the latent space of the neural networks revealed that all four model were able to differentiate between MS and the other categories. For the IVAE-256 model we achieved a precision value of 92\% and recall of 89\% for detecting MS.
To determine whether our models were really picking up on disease specific details of the image we searched for biases in the different datasets within the training database that could lead to differences in the images. We found such biases, it was not possible to assert with certitude if they had an impact on the classification result.

\bibliographystyle{plain}
\bibliography{ref}
\end{document}